\documentclass[aps,pra,reprint,groupedaddress]{revtex4-1}
\usepackage[utf8]{inputenc}
\usepackage{amsmath}
\usepackage{amssymb}
\usepackage{amsthm}
\usepackage{mathtools} 
\usepackage{physics}   
\usepackage{bm}        
\usepackage{dsfont}    
\usepackage{mathrsfs}  
\usepackage{wasysym}   
\usepackage{graphicx}
\usepackage{xcolor}
\usepackage{float}
\usepackage[normalem]{ulem}
\usepackage{soul}
\usepackage{verbatim}
\usepackage{tabularray}
\usepackage{appendix}
\usepackage{chngcntr}
\usepackage{hyperref}
\usepackage{cleveref}

\hypersetup{
    colorlinks=true,
    linkcolor={red!50!black},
    citecolor={blue!50!black},
    urlcolor={blue!80!black}
}

\counterwithout{table}{section}

\newcommand{\beq}{\begin{equation}}
\newcommand{\edq}{\end{equation}}
\newcommand{\bes}{\begin{subequations}}
\newcommand{\eds}{\end{subequations}}
\newcommand{\beqn}{\begin{equation*}}
\newcommand{\edqn}{\end{equation*}}

\newcommand{\an}[1]{\textcolor{black}{#1}}

\begin{document}
\title{Non-Markovianity and memory enhancement in Quantum Reservoir Computing}

\author{Antonio Sannia}
\email{sannia@ifisc.uib-csic.es}
\affiliation{%
 Institute for Cross-Disciplinary Physics and Complex Systems (IFISC) UIB-CSIC, Campus Universitat Illes Balears, 07122, Palma de Mallorca, Spain.
}%

\author{Ricard Ravell Rodríguez}%
\affiliation{%
 Institute for Cross-Disciplinary Physics and Complex Systems (IFISC) UIB-CSIC, Campus Universitat Illes Balears, 07122, Palma de Mallorca, Spain.
}%

\author{Gian Luca Giorgi}
\affiliation{%
 Institute for Cross-Disciplinary Physics and Complex Systems (IFISC) UIB-CSIC, Campus Universitat Illes Balears, 07122, Palma de Mallorca, Spain.
}%

\author{Roberta Zambrini}%
\affiliation{%
 Institute for Cross-Disciplinary Physics and Complex Systems (IFISC) UIB-CSIC, Campus Universitat Illes Balears, 07122, Palma de Mallorca, Spain.
}%

%\date{\today}
\begin{abstract}
    Featuring memory of past inputs is a fundamental requirement for machine learning models processing time-dependent data. In quantum reservoir computing, all architectures proposed so far rely on Markovian dynamics, which, as we prove, inherently lead to an exponential decay of past information, thereby limiting long-term memory capabilities. We demonstrate that non-Markovian dynamics can overcome this limitation, enabling extended memory retention. By analytically deriving memory bounds and supporting our findings with numerical simulations, we show that non-Markovian reservoirs can outperform their Markovian counterparts, particularly in tasks that require a coexistence of short- and long-term correlations. We introduce an embedding approach that allows a controlled transition from Markovian to non-Markovian evolution, providing a path for practical implementations. Our results establish quantum non-Markovianity as a key resource for enhancing memory in quantum machine learning architectures, with broad implications in quantum neural networks.
\end{abstract}

\maketitle

\section{Introduction}

Within the field of machine learning, there is growing interest in neural networks capable of processing time-dependent data presenting long-range correlations ~\cite{Lim2021}. Architectures such as long short-term memory~\cite{LSTM} and transformers~\cite{vaswani2023attentionneed} are currently producing outstanding results in time series forecasting and language processing. However, the high energy cost of these deep learning models is a significant limiting factor for their scalability~\cite{Desislavov2023,Samsi2023,anthony2020,park2018}. As a promising alternative, analog implementations of machine learning models offer a potential solution to this challenge~\cite{Markovi2020,Wright2022,Hu2018}.

In this context, Reservoir Computing~\cite{Nakajima2020,RC_book} has attracted significant attention due to its versatility in handling both static and temporal tasks, its training efficiency (requiring only linear regression), and its adaptability to diverse physical platforms without the need for fine-tuning of experimental parameters~\cite{Cucchi2022}. Among the various proposals \cite{RC_book}, quantum systems allow the exploration of unique possibilities such as having an exponential number of degrees of freedom and direct embedding of quantum inputs without the need for tomography~\cite{Fujii2017,Mujal2021}. Several works have been done to connect the performances of quantum reservoir computing (QRC) to physical properties such as dynamical phase transitions~\cite{MartnezPea2021,Res_Probing}, quantum dissipation~\cite{Sannia2024, Optical_Abs, Cheamsawat2025}, quantum chaos~\cite{Llodr2025}, squeezing~\cite{GarcaBeni2024}, topological effects~\cite{Sannia2025}, entanglement~\cite{Gtting2023}, quantum back-action~\cite{Mujal2023,back_act} and quantum coherences~\cite{Palacios2024}.

One of the key features of QRC is the presence of an internal memory, which enables the processing of temporal information, a critical capability for tasks such as time series analysis and forecasting~\cite{Mujal2021}.
The specific memory featured by the reservoir in its dynamics is therefore a fundamental resource for QRC, and a complete characterization of it is still missing. In the following, we address this fundamental question by showing that all reservoir models proposed so far, relying on a Markovian evolution, inevitably erase information about past inputs at an exponential rate. In fact, while Markovian reservoir models enable the inference of unlimited data streams and allow for memory adjustment~\cite{Sannia2024,Hu2024}, their exponential memory decay fundamentally restricts their applicability. Importantly, we identify quantum non-Markovianity as the key mechanism to overcome this limitation, opening the possibility of studying problems where longer memory is needed. 

Quantum non-Markovianity refers to the phenomenon in which the evolution of an open quantum system shows significant memory effects, so that its future behavior depends not only on its current state but also on its past history~\cite{Breuer2016,rivas2012open,devega2017dynamics,CHRUSCINSKI20221,Liu2011}. This can lead to revivals of coherence, entanglement, and other quantum properties that would otherwise decay ~\cite{Breuer2009,xu,wang,buscemi}. The mathematical characterization of various aspects of non-Markovianity has been an active area of research in recent years, and its role as a resource in quantum technologies has been shown to be highly context-dependent. It has been shown to be beneficial in quantum metrology~\cite{Alex2012}, control~\cite{Reich2015}, teleportation~\cite{Laine2014}, information processing~\cite{Bylicka2014}, computing~\cite{Dong2018}, and entanglement generation~\cite{Thorwart_2009}. However, it can also be detrimental in certain scenarios, such as inhibiting the emergence of synchronization~\cite{Karpat_2021} and negatively impacting specific classes of quantum algorithms~\cite{Rossi2018}. 

In this work, we demonstrate how non-Markovianity can enhance memory and, consequently, the performance of QRC. Specifically, we provide theoretical justifications and support our claims with both analytical and numerical results. Additionally, we propose potential implementations where non-Markovianity can be introduced and controlled through an embedding method~\cite{devega2017dynamics}. \an{This newly introduced framework exploits a physical memory effect intrinsic to open quantum systems, a resource that has remained unexploited in previous reservoir computing architectures. By harnessing the information backflow that characterizes non-Markovian dynamics, our approach enables the reservoir to retain correlations with past inputs over extended timescales, overcoming the fundamental exponential memory decay inherent to Markovian quantum reservoirs. We demonstrate this advantage through two complementary implementations: first, an analytically tractable Quantum Residual Reservoir model that exhibits controlled memory revivals at prescribed delays (Sect.~\ref{Sec:QRS}), and second, an experimentally realizable embedded non-Markovian reservoir where tunable depolarizing channels allow seamless interpolation between Markovian and non-Markovian regimes  (Sect.~\ref{Sec:EMB}).}

\begin{figure}[t!]
    \centering  \includegraphics[width=\linewidth, keepaspectratio]{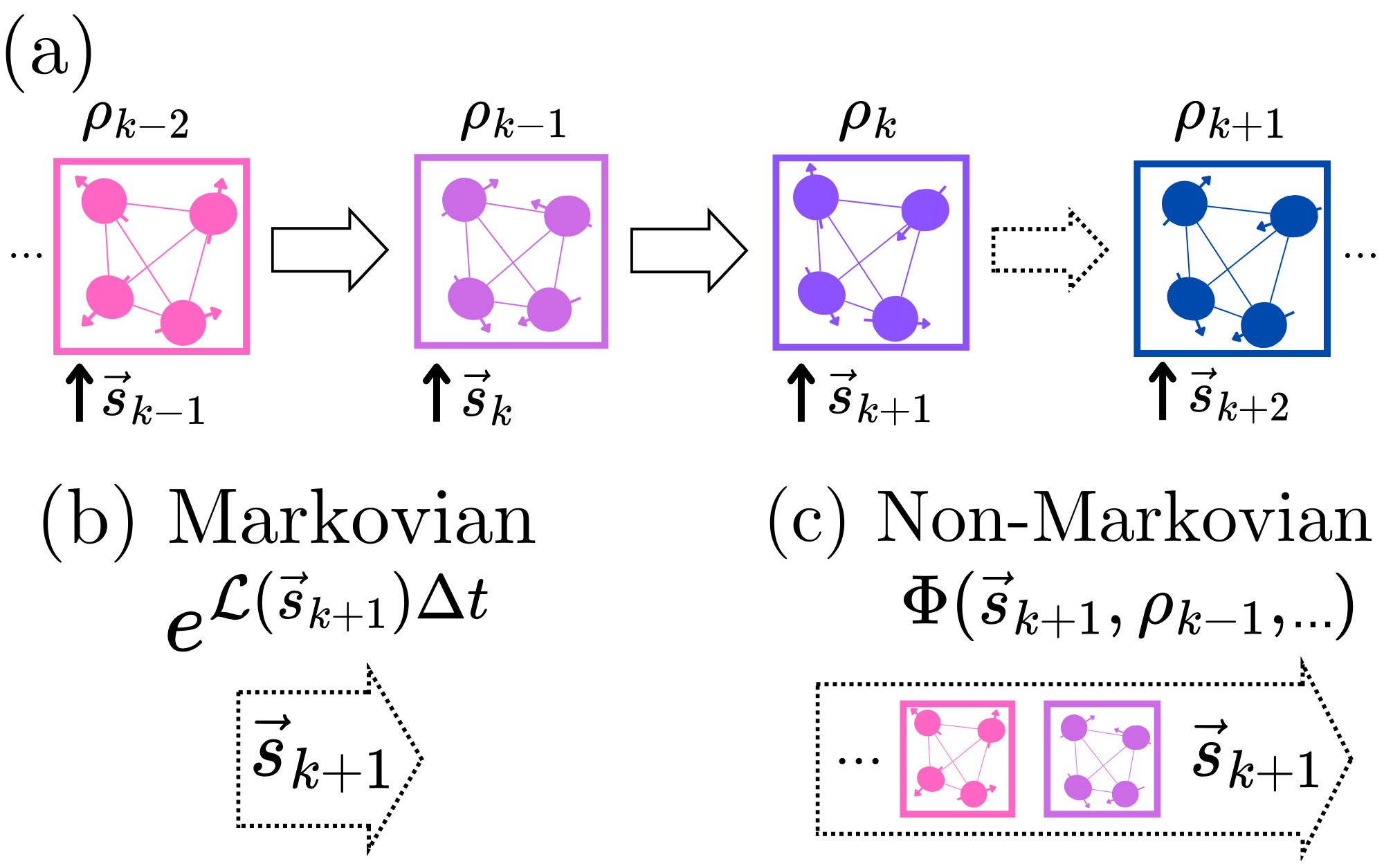}
\caption{Scheme of a reservoir evolution (a). In the Markovian case (b), the reservoir state updating rule depends solely on the injected input, whereas in the non-Markovian one (c), past states also play a role.}    \label{fig:scheme}
\end{figure}

\section{Memory bounds of Markovian reservoirs} 
The Gorini-Kossakowski-Sudarshan-Lindblad (GKLS) Master Equation~\cite{gorini1976completely,lindblad1976generators}
\begin{equation*}
    \frac{d\rho}{dt} = \mathcal{L}[\rho] = -i [H, \rho] + \sum_i \gamma_i \left( L_i \rho L_i^\dagger - \frac{1}{2} \{L_i^\dagger L_i, \rho\} \right),
\end{equation*}
is known to describe the Markovian evolution of an open system
with Hamiltonian $H$, where $\{L_i\}$ are the jump operators describing the interaction with the external environment, and $\{\gamma_i\}$ are the decay rates. 
In QRC, the reservoir (an open quantum system) is driven by an input sequence, while a set of measured observables form the output, in a three-layer configuration (see \cite{Supp,Mujal2021}). Given time series of input vectors $\{\dots,\vec{s}_{k-1},\vec{s}_{k},\vec{s}_{k+1},\dots\}$, where $k$ iterates over the time-steps,  a Markovian quantum reservoir state evolves according to the rule:
\begin{equation}\label{Eq:Mark_model}
    \rho_{k+1} = e^{\mathcal{L}(\vec{s}_{k+1}) \Delta t} \rho_{k}
\end{equation}
where $\mathcal{L}(\vec{s}_{k+1})$ is a parametric Liouvillian that defines the particular reservoir model and $\Delta t$ is the evolution time, which can be considered as a model hyperparameter (see Fig.~\ref{fig:scheme} (b)).

After each input injection, the reservoir response is optimized for the target task, with a simple linear regression  of observable expectation values. In particular, considering a set of observables $\{\mathcal{O}_i\}$, the reservoir output $y_k$ is given by the expression:
\begin{equation}\label{Eq:out}
    y_k = \omega_{0} + \sum_i \omega_i \Tr{\rho_k \mathcal{O}_i}, 
\end{equation}
where the coefficients $\omega_i$ are optimized, in a supervised way, depending on the problem of interest~\cite{Cucchi2022}.

For the proper operation of a reservoir computing model, different properties need to be satisfied, such as the echo state property~\cite{Yildiz2012}. This means that the reservoir will produce the same outcome independently of the initial state if the same input series is injected. The echo state property holds for any input sequence if the stationary state $\mathcal{L}(\vec{s}_{k})$ is unique for every input $\vec{s}_{k}$~\cite{Sannia2024}. The converse implication is straightforward to verify. In the following, we will restrict to Liouvilians satisfying this condition. Moreover, if $\mathcal{L}(\vec{s}_{k})$ is a continuous function of $\vec{s}_{k}$, then the reservoir state will depend only on the recent input history, satisfying the so-called fading memory~\cite{Sannia2024}. Under the fading memory property, each output $y_k$ can be considered %to be 
a time-invariant function and, consequently, can be expanded according to a Volterra series~\cite{Boyd1985}. In our notation, the Volterra expansion takes the following form:
\begin{align}\label{Eq:Volt_yk}
   y_k = \sum_{n = 1}^{\infty} \prod_{\tau_1 = 0}^{\infty}  \prod_{\tau_2 = \tau_1}^{\infty} \cdots \prod_{\tau_n = \tau_{n-1}}^{\infty} \mathcal{P}_1(\vec{s}_{k-\tau_1}) \dots \mathcal{P}_n(\vec{s}_{k-\tau_n}) \nonumber \\
    = \sum_{n = 1}^{\infty} f_n(\vec{s}_{k-\tau_1}, \dots, \vec{s}_{k-\tau_n})
\end{align}
where $\mathcal{P}_i(\vec{s}_{k-\tau_i})$ are polynomials of degree one with respect to the $\vec{s}_{k-\tau_i}$ entries, and $f_n$ are functionals that group all the monomials in the series that have input dependence up to a delay $\tau_{n}$.

Given a Markovian QRC with reservoir dynamics governed by Eq.~\eqref{Eq:Mark_model}, we show in the Supplemental material (see Ref.~\cite{Supp}) that for each element of the output Volterra expansion% it holds that:
\begin{equation}\label{Eq:decay}
 |f_n(\vec{s}_{k-\tau_1}, \dots, \vec{s}_{k-\tau_n})| = \mathcal{O}(e^{-a\tau_n}),  \end{equation}
where $a$ is a positive constant that depends on the particular reservoir model.
Importantly, Eq.~\eqref{Eq:decay} implies fundamental bounds on the reservoir memory. Considering a $T$ steps evolution,  we quantify the reservoir's ability to reproduce past input functionals with the capacity introduced in Ref.~\cite{Dambre2012}:
\begin{equation}\label{Eq:capacity}
    C[\hat{y}] = 1 - \min_{\{\omega_i\}}\frac{\textit{MSE}_T[\hat{y}]}{\langle y^2 \rangle_T},
\end{equation}
where $\{\hat{y}_k\}_{k=1}^T$ is the target series; \textit{MSE} indicates the mean square error: $\textit{MSE}_T = \sum_{k=1}^T(\hat{y}_k-y_k)^2/T$ (depending on the weights in Eq.~\eqref{Eq:out}); and $\langle y^2 \rangle_T = \sum_{k=1}^T y_k^2/T$. 

Among the properties of the capacity \cite{Dambre2012}, we recall that $0 \leq C \leq 1$, where the extreme cases $C=0$ and $C=1$ correspond to a complete mismatch between the target and predictions, and to a perfect agreement, respectively. Combining now Eqs.~\eqref{Eq:decay} and~\eqref{Eq:capacity}, and considering a target $\hat{y}(\tau)$ that is a function only of past inputs delayed by a certain $\tau$ (e.g. $\hat{y}_k(\tau) = \vec{s}_{k-\tau}^{\,2}$) we find the following capacity scaling:  
\begin{equation}\label{Eq:cap_scal}  
C[\hat{y}(\tau)] = \mathcal{O}(e^{-a\tau}).  
\end{equation}
The interpretation of Eq.~\eqref{Eq:cap_scal} is that a Markovian reservoir tends to \textit{exponentially} erase information of past inputs over time. This conclusion can be directly generalized in the case in which the inputs, instead of being vectors of classical parameters, are quantum states, making the presented considerations valid even for quantum tasks.

As we will show below, for some kinds of tasks, a longer memory is required, and the introduction of quantum non-Markovianity is necessary for overcoming this exponential decay. 
In particular, we introduce a more general class of quantum reservoir computers whose updating rule is
\begin{equation}\label{Eq:nonmark_res}
    \rho_{k+1} = \Phi (\vec{s}_{k+1}, \rho_{k-1},\cdots ) \rho_{k},
\end{equation}
where $\Phi$ is a super-operator corresponding to a non-Markovian physical process in which an external environment has a memory of the previous reservoir states (see Fig.~\ref{fig:scheme} (c)). \an{In general, the resulting family of dynamical maps is not CP-divisible across the discrete algorithmic time steps, since $\Phi$ cannot be defined independently of earlier reservoir states.}

\begin{figure}[t!]
    \centering  \includegraphics[width=\linewidth, keepaspectratio]{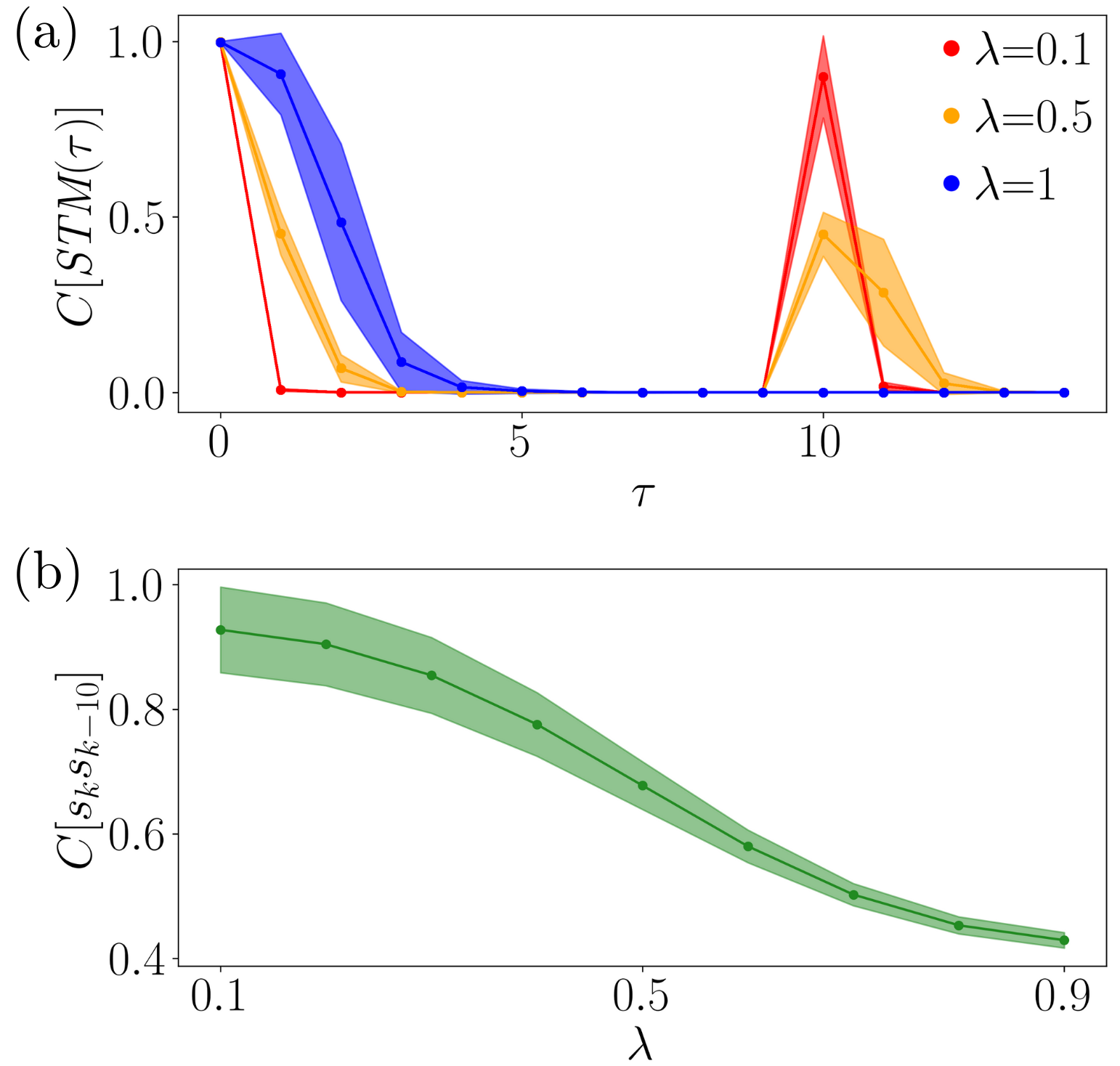}
\caption{Quantum Residual Reservoir performance analysis. (a) Capacity for the short-term memory task, as a function of the delay $\tau$. (b) Capacity of reproducing the monomial $s_ks_{k-10}$, varying $\lambda$ in steps of $0.1$. All the capacity values have been averaged over $100$ random realizations of reservoir initial states, Hamiltonian couplings, and input sequences~\cite{Supp}. The hyperparameters have been fixed to be $N=3$, $\Delta t = 10$, $h=1$ and $\gamma = 0.1$.}    \label{fig:rs}
\end{figure}

\section{Quantum Residual Reservoir}\label{Sec:QRS} We now give an illustrative example of a non-Markovian process that allows us to overcome the exponential scaling previously found. For generic Markovian reservoir, defined from a Liouvillian $\mathcal{L}(\vec{s}_{k})$, we can always define a fading memory time scale $\tau_{FM}$ such that all the delayed inputs $\vec{s}_{k-\tau}$, with $\tau > \tau_{FM}$, are practically irretrievable from the reservoir state $\rho_k$ \cite{Sannia2024,Mujal2023}. Let us consider instead the following non-Markovian QRC evolution:
\begin{equation}\label{Eq:rs_mod}
\rho_{k+1} = e^{\mathcal{L}(\vec{s}_{k+1}) \Delta t} [ \lambda \rho_{k} + (1-\lambda)\rho_{k-\tau_{E}} ],
\end{equation}
where $\lambda$ is a model hyperparameter that takes values in the interval $[0,1)$, and $\tau_{E}$ is the non-Markovian environment time scale, which we consider to be larger than $\tau_{FM}$. Interestingly, Eq.~\eqref{Eq:rs_mod} represents a quantum counterpart of the recently introduced Residual reservoir computers~\cite{Ceni2024}, making use of dilated skip connections~\cite{chang2017dilated}.

For this non-Markovian case, it is now immediate to see that the state $\rho_{k}$ is statistically correlated with the past inputs following an exponential decay trend, inside the fading memory window of $\mathcal{L}(\vec{s_k})$, meanwhile, for inputs delayed from a factor at least equal to $\tau_{E}$, we find an information retrieval that overcomes the fading memory time. Importantly, the parameter $\lambda$ governs the interplay between these competing behaviors --the Markovian loss of input information and the non-Markovian revival--with its optimal value being task-dependent.

To numerically verify the memory properties of this Quantum Residual Reservoir, we start from the Markovian model introduced in Ref.~\cite{Sannia2024}. Specifically, considering the input sequence to be composed of scalar values $s_k=[0,1]$, the parametric Liouvillian will act according to the equation
\begin{equation}\label{eq:ME}
    \mathcal{L}(s_k)[\rho] = -i[H(s_k),\rho] + \mathcal{D}_L[\rho],
\end{equation}
where $H(s_k)$ is an input-dependent Hamiltonian that reads 
\begin{equation}\label{eq:HI}
H=\sum_{i<j}^{N}J_{ij} \sigma_{i}^{x}\sigma_{j}^{x}+h\sum_{i=1}^{N}\sigma_{i}^{z} + h(s_k+1) \sum_{i=1}^N \sigma^x_i,
\end{equation}
where $N$ is the number of reservoir qubits, $J_{ij}$ are the random couplings, sampled from the fixed interval $[-1,1]$, $h$ is the strength of the magnetic fields, and $\mathcal{D}_L$ is a local dissipator:
\begin{equation*}
    {\cal D}_{L}[\rho]\equiv\gamma \sum_{i=1}^N ( \sigma_i^{-}\rho  \sigma_i^{+}-\frac{1}{2} \{ \sigma_i^{+}\sigma_i^{-},\rho \})
\end{equation*}
where $\gamma$ is its corresponding decay rate.

We can now study the Quantum Residual Reservoir's ability to retrieve past inputs by computing the performance for the short-term memory task (STM)~\cite{Fette2005}. According to the standard definition, at each time step, the target values $\hat{y}_k$ are the past inputs delayed by a time $\tau$: $\hat{y}_k = s_{k-\tau}$, \an{while the injected inputs are i.i.d. uniform random variables over the interval $[0,1]$}. Fixing $\tau_E = 10$ and considering the output observables to be the single-qubit matrices $\{\sigma^z_i\}_{i=1}^N$, in Fig.~\ref{fig:rs}~(a) we plot the estimated capacity as a function of the delay $\tau$ for different values of $\lambda$ \cite{Supp}. As predicted by the theoretical results, when the evolution is Markovian ($\lambda = 1$), the capacity exhibits a typical decay, making inputs delayed by $\tau > 3$ practically irretrievable. Moreover, as expected in the non-Markovian cases ($\lambda = 0.5$ and $\lambda = 0.1$), we observed a capacity revival at $\tau = \tau_E$, whose magnitude increases as $\lambda$ decreases.

This memory revival, observed exclusively in non-Markovian dynamics, can be crucial for solving certain tasks. For instance, Fig.~\ref{fig:rs}~(b) illustrates that the Quantum Residual Reservoir’s capacity for the target $\hat{y}_k = s_k s_{k-10}$ tends to decrease with $\lambda$~\cite{Supp}. This behavior suggests that the task requires the reservoir to retain both recent inputs ($s_k$) and long-delayed ones ($s_{k-10}$). Achieving this dual time-scale memory is only possible by introducing non-Markovianity. While in this synthetic QRC design, the performance in the chosen non-linear memory task is monotonically related to the revival strength, in general, the form of dissipation and non-Markovianity cannot be isolated, and moreover different memory forms are required in different tasks. Therefore, the enhancing effect of non-Markovianity will depend on the QRC task.

\section{Embedded Non-Markovian reservoir}\label{Sec:EMB} 

\begin{figure}[t!]
    \centering  \includegraphics[width=\linewidth, keepaspectratio]{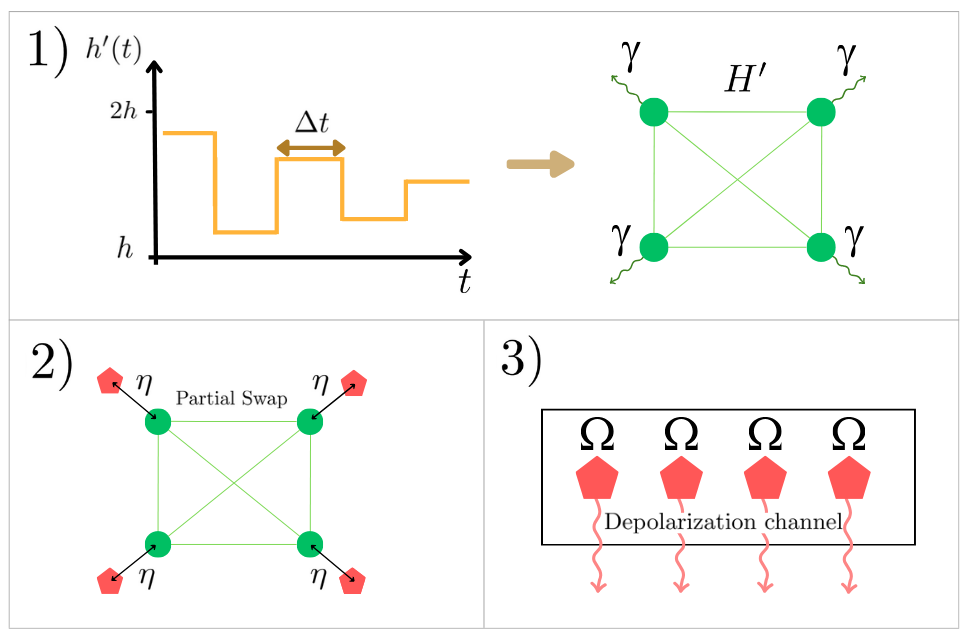}
\caption{Schematic representation of the main steps in the evolution rule of the embedded non-Markovian model presented in the main text: (1) Input is injected through Hamiltonian driving in the Lindbladian; (2) interaction with auxiliary qubits occurs via partial swaps; (3) depolarizing channels act on the auxiliary qubits to tune the degree of non-Markovianity.}    \label{fig:scheme_emb}
\end{figure}

Non-Markovian dynamics of open quantum systems can be obtained from a Markovian evolution in a larger space, tracing out some degrees of freedoms~\cite{devega2017dynamics,Bay1997,Imamoglu1994,Garraway1997,Garraway1996,Breuer2004,Arrigoni2013,Dorda2014}. We will follow this approach, known as embedding method~\cite{devega2017dynamics}, for assessing the memory and forecasting of QRC tuned from a Markovian to non-Markovian operation. We consider as a reservoir the open quantum system model of Ref.~\cite{Rijavec2025}, whose evolution rule can be factorized in the following form:
\begin{equation*}
    \rho_{k+1} = \Lambda \circ e^{\mathcal{L}(s_k)\Delta t}[\rho_k],
\end{equation*}
where $\mathcal{L}(s_k)$ is the same Liouvillian of Eq.~\eqref{eq:ME} and $\Lambda$ is a non-Markovian channel that involves the presence of auxiliary qubits. In particular, given the quantum reservoir in Eq.~\eqref{eq:HI}, we couple each of the $N$ qubits to an auxiliary one to implement the $\Lambda$ action \an{(see Figure~\ref{fig:scheme_emb} for a schematic summary of the model)}. First of all, at each time step $k$, every auxiliary–reservoir qubit pair interacts with a partial Swap operation
\begin{equation*}
    \hat{P}_i(\eta)=\cos (\eta) \mathbb{I}+i\sin (\eta)\hat{\mathcal{S}}_i
\end{equation*}
where $i$ is an index ranging from $1$ to $N$; $\mathbb{I}$ represents the identity operator over the entire space generated by the $2N$ qubits; $\hat{\mathcal{S}}_i$ denotes the Swap operator acting on the $i$th qubits of the reservoir and the auxiliary system; and $\eta$ is a hyperparameter in the set $[0, \pi/2)$, interpreted as the interaction strength between the system and the auxiliary qubits. After the application of all the partial Swaps, a set of depolarizing channels acts on the auxiliary system: 
\begin{equation*}
\Delta_{\Omega}^i[\rho] =\sum_{m=0}^3 \hat{K}_i^m \rho \hat{K}_i^{m\dagger},
\end{equation*}
where $\hat{K}_i^m$ are the Kraus operators acting on the $i$-th auxiliary qubit: $\hat{K}_i^0 \coloneqq \sqrt{1-3\Omega/4}\,\mathbb{I}_i$, $\hat{K}_i^1 \coloneqq \sqrt{\Omega/4}\,\sigma^x_i$, $\hat{K}_i^2 \coloneqq \sqrt{\Omega/4}\,\sigma^y_i$, $\hat{K}_i^3 \coloneqq \sqrt{\Omega/4}\,\sigma^z_i$, and $0\leq\Omega\leq 1$. 

Importantly, $\Omega$ represents the strength of the depolarizing channels and allows tuning the non-Markovianity level. In particular, in the case of $\Omega=1$, the reservoir dynamics is fully Markovian because the auxiliary system, at each time step, will be prepared to a fully mixed state, having no memory of past reservoir states. On the other hand, as shown in Ref.~\cite{Rijavec2025}, it is expected that decreasing the $\Omega$ value corresponds to an increase of non-Markovianity, up to the maximum case for $\Omega = 0$, where the partial Swap is applied with auxiliary qubits in the state of the previous step leading to the maximum back-flow of memory in the system. In the supplementary material, we show a numerical verification of this behavior~\cite{Supp}.

\begin{figure}[t]
    \centering  
    \includegraphics[width= \linewidth, keepaspectratio]{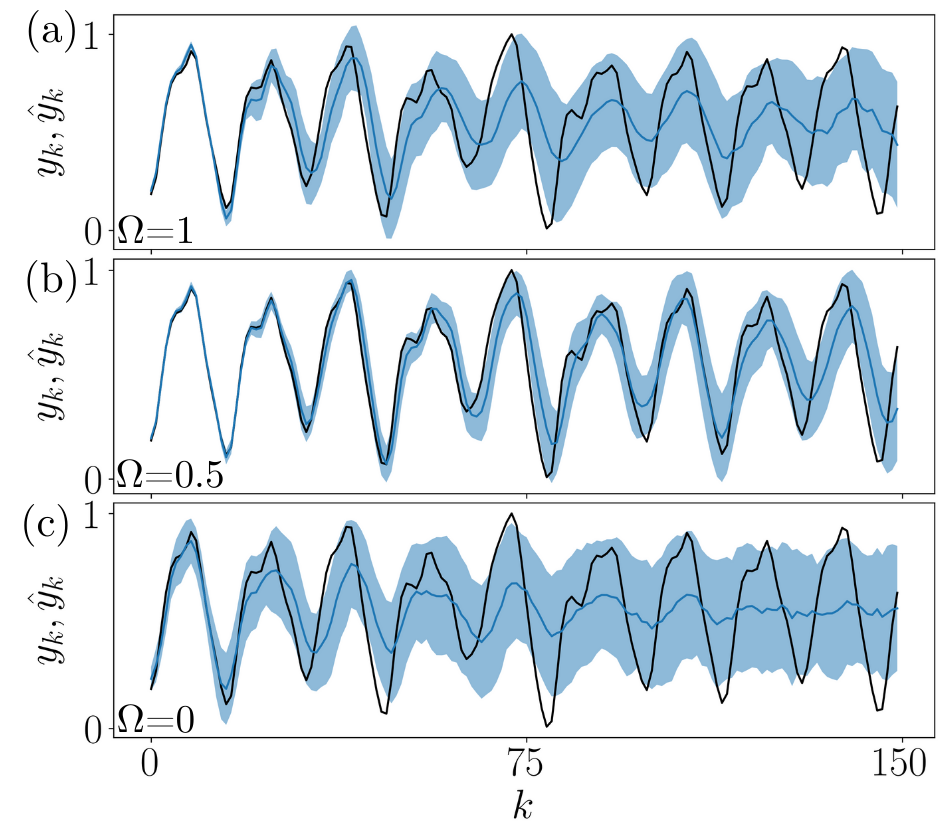}
\caption{Reservoir predictions of the Mackey-Glass time-series during a test phase of $150$ points. The black lines correspond to the numerical solutions of the Mackey-Glass equation, while the blue lines are the average of predictions over $100$ random realizations of reservoir initial states and Hamiltonian couplings~\cite{Supp}. The blue shadow regions are the statistical errors, chosen as one standard deviation. Different $\Omega$ values have been considered for changing the non-Markovianity amount: (a) $\Omega=1$, (b) $\Omega = 0.5$, (c) $\Omega = 0$. The other hyperparameters have been fixed to the following values: $h=1$, $dt=0.5$, $\gamma=0.1$, $\eta = \pi/4$, and $2N = 8$. The performances have been quantified through the average mean square error between the numerical solution of the Mackey-Glass equation and the reservoir predictions, taking $150$ points in the test phase. The values found, in descending order with respect to $\Omega$, are respectively $1.8 \cdot 10^{-2}$, $6.8 \cdot 10^{-3}$, and $2 \cdot 10^{-2}$.}\label{Fig:Emb}
\end{figure}

Turning now to QRC tasks, we examine how non-Markovianity influences the reservoir capabilities of chaotic time series forecasting. Specifically, we focus on the widely studied Mackey-Glass (MG) series~\cite{Mackey1977}, where the target values satisfy the delayed differential equation:  \begin{equation}\label{eq:MG}  
    \dot{s}(t) = -0.1s(t) + \frac{0.2s(t-\tau)}{1 + s(t-\tau)^{10}},  
\end{equation}  
and we set $\tau = 17$ to ensure a chaotic regime~\cite{farmer1987predicting,jaeger2004harnessing}.
During the training phase, we fed the system with the numerical solution of Eq.~\eqref{eq:MG}, solved with a time resolution of $t_r = 3$ (see Ref.~\cite{Ortn2015} for details), with input values rescaled to the interval $[0,1]$. The output coefficients are trained to perform a one-step-ahead prediction task, $\hat{y}_k = s_{k+1}$~\cite{Supp}, using measured observables that correspond to Pauli strings of length one ($\sigma_i^a$) and two ($\sigma^a_i \sigma^b_j$), where $a,b \in \{x,y,z\}$ and $1 \leq i,j \leq N$ with $i \neq j$. Finally, during the evaluation phase, the system evolved autonomously, using its previous output as the current input at each time step. 

\begin{figure}[t!]
    \centering  \includegraphics[width=0.85\linewidth, keepaspectratio]{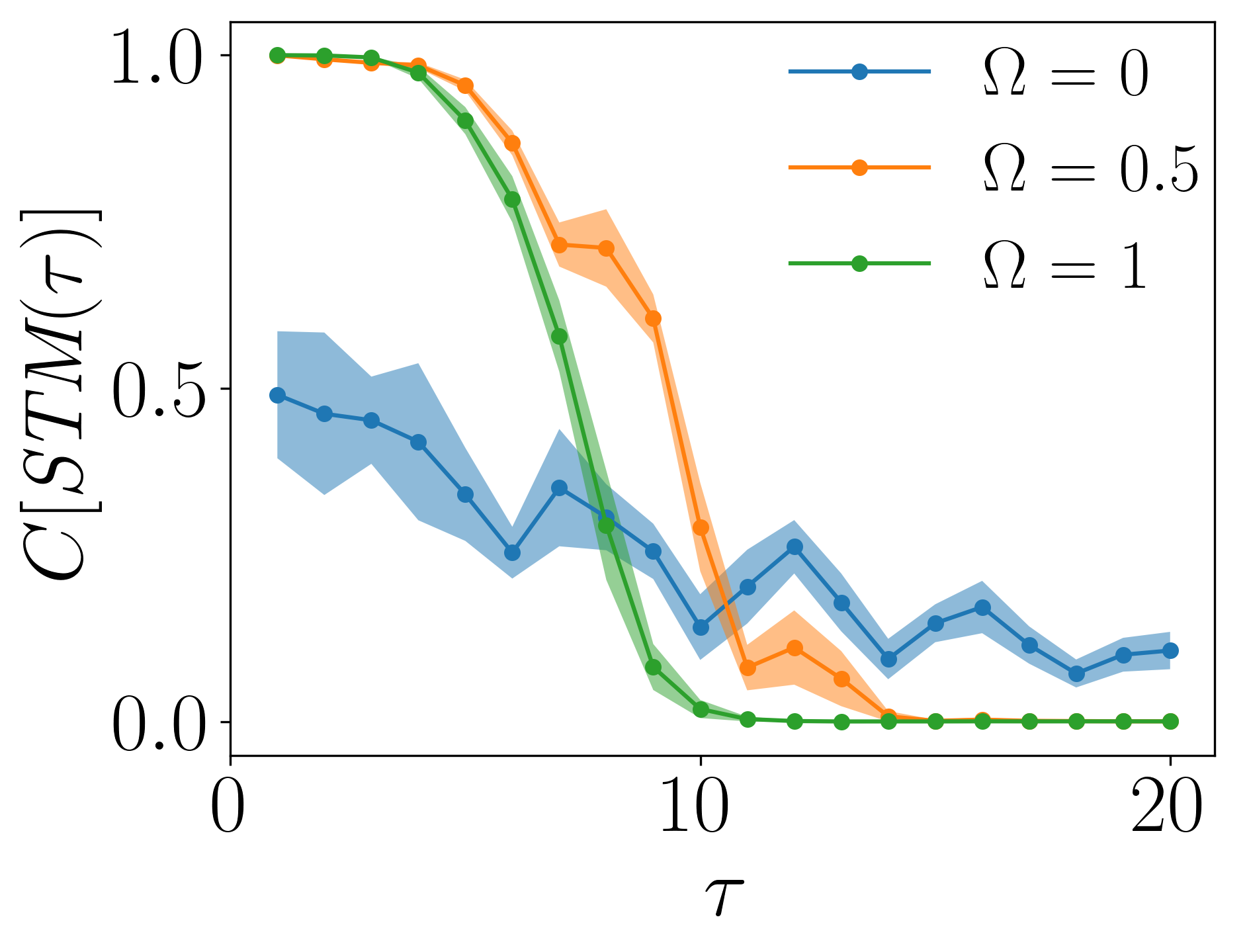}
\caption{Capacity of the short-term memory task as a function of the delay $\tau$, considering different $\Omega$ values. The shadow regions cover one standard deviation calculated randomly varying the Hamiltonian, the initial reservoir state, and the injected sequence. According to the main text, the other hyperparameters have been fixed to: $h=1$, $dt=0.5$, $\gamma=0.1$, $\eta = \pi/4$ and $2N = 8$.} \label{fig:capacity_emb}
\end{figure}

In Fig.~\ref{Fig:Emb}, we present the reservoir predictions during the test phase for different values of $\Omega$. The results clearly indicate that the best forecast occurs in the non-Markovian regime with $\Omega = 0.5$, outperforming the fully Markovian one obtained for $\Omega=1$. Moreover, as expected, the degree of non-Markovianity in the updating rule must be tuned to achieve an advantage. Notably, when $\Omega = 0$, corresponding to the maximum amount of non-Markovianity, the reservoir's forecasting ability deteriorates. \an{An explanation of this behavior can be obtained by evaluating the short-term memory task for the embedded model. From Fig.~\ref{fig:capacity_emb}, we observe that the Markovian case ($\Omega=1$) leads to a monotonic capacity decay, in agreement with our theoretical results. In contrast, the non-Markovian regimes ($\Omega=0$, $\Omega=0.5$) exhibit memory revivals. Notably, in the extreme non-Markovian limit ($\Omega=0$), although revivals are present, the reservoir displays only weak correlations with past inputs overall. This indicates that memory contributions become broadly distributed but individually small, which ultimately deteriorates performance in demanding tasks such as Mackey-Glass time-series prediction.}

As anticipated, the potential performance gains from introducing non-Markovianity depend on the specific task at hand. Looking instead at a different forecasting task, non-Markovianity does not yield any improvement for the Santa Fe laser task~\cite{Weigend} (see supplementary material~\cite{Supp}). One possible explanation is that, in this task, the system is not intended to evolve autonomously, as in the Mackey-Glass task, as the reservoir is continuously driven by the true series values, potentially making the prediction problem inherently easier.

\an{Moreover, the proposed non-Markovian embedding is not limited to the specific interaction considered here (the partial Swap unitary) or to the use of depolarizing channels. Alternative realizations can be constructed using experimentally accessible quantum gates, for example entangling operations~\cite{Hu2024,Senanian2024,Chen2020}, or by adopting other physically relevant noise models such as amplitude damping~\cite{Sannia2024,Verstraete2009}. Importantly, the mechanism does not rely on fine tuning of a particular gate parameter: the advantage arises from the presence of a structured information exchange between the reservoir and the auxiliary degrees of freedom. Moderate imperfections in the implemented interactions, like for example errors in the partial swaps angles, are therefore not expected to qualitatively suppress the effect. The size and structure of the auxiliary environment can be adjusted independently of the reservoir qubits, enabling the design of multiple time scales and memory effects. In addition, the auxiliary qubits are never measured, so the measurement overhead remains the same as in a standard $N$-qubit reservoir.}

\an{This embedded strategy differs fundamentally from simply increasing the size of a purely Markovian reservoir. Although enlarging a Markovian reservoir can extend memory, it does not alter the characteristic exponential decay of Markovian dynamics. In contrast, our embedding introduces controlled information backflow and modifies temporal correlations. For tasks that require correlations across multiple time scales, this change in the dynamics can be more beneficial than a mere increase in system size.}

\an{Finally, the proposed framework is not limited to the specific mechanisms introduced here. Any intrinsic non-Markovianity present in a given quantum platform can serve as an additional resource to enhance memory retention. This versatility makes our approach widely applicable across different physical implementations, from superconducting circuits to trapped ions and photonic systems, and readily adaptable to the memory requirements of diverse temporal tasks.}

\vspace{1em}

\section{Conclusions} 

We have established fundamental limitations on memory capacity in Markovian QRC and demonstrated how non-Markovianity can enhance memory retention. Our results show that Markovian reservoirs inevitably experience an exponential decay of past memory, restricting their information capacity for long-term dependencies. By incorporating non-Markovian dynamics in an illustrative example, we have demonstrated an extension of the reservoir’s memory through a revival mechanism and the enhancing effect in a non-linear memory task requiring long-range correlations. While in this case the performance improvement is monotonous in the revival strength, this is not the common situation. With an embedding approach  enabling the controlled introduction of non-Markovianity in QRC, we showed that non-Markovian reservoirs can outperform their Markovian counterparts in chaotic time series forecasting, illustrating the practical benefits of memory effects in quantum machine learning. Our results establish non-Markovianity as a resource for QRC, with broader implications in quantum information processing. \an{Crucially, the memory enhancement demonstrated here is not achieved through classical means such as network enlargement or delayed feedback, but through a physical mechanism intrinsic to open quantum dynamics, underscoring the unique computational advantage offered by non-Markovian quantum systems.} An open, challenging question is to establish a comprehensive theoretical framework in QRC 
to relate reservoirs' memory and non-Markovianity to their performance on different temporal tasks.

\section*{Acknowledgements}

We acknowledge the Spanish State Research Agency, through the María de Maeztu project CEX2021-001164-M funded by the MICIU/AEI/10.13039/501100011033, through the COQUSY project PID2022-140506NB-C21 and -C22 funded by MICIU/AEI/10.13039/501100011033 and by ERDF, EU, MINECO through the QUANTUM SPAIN project, and EU through the RTRP— NextGenerationEU within the framework of the Digital Spain 2025 Agenda. The CSIC Interdisciplinary Thematic Platform (PTI+) on Quantum Technologies in Spain (QTEP+) is also acknowledged. The project that gave rise to these results received the support of a fellowship from the “la Caixa” Foundation (ID 100010434). The fellowship code is LCF/BQ/DI23/11990081.

\bibliographystyle{bibstyle}
\bibliography{refs.bib}

\section*{Author Contributions} 
A.S. performed the analytical calculations. A.S. and R.R.R. carried out the numerical calculations. A.S., R.R.R., G.L.G. and R.Z. analysed the results. A.S., R.R.R., G.L.G. and R.Z. wrote the manuscript.

\section*{Corresponding Author}
Correspondence to Antonio Sannia.

\section*{Competing interest}
The authors declare no competing interests.

\section*{Funding}
Spanish State Research Agency: A.S, R.R.R., G.L.G. and R.Z. Spanish Ministry of Economic Affairs and Digital Transformation: A.S, R.R.R., G.L.G. and R.Z. Consejo Superior de Investigaciones Científicas: A.S, R.R.R., G.L.G. and R.Z. "La Caixa" Foundation: A.S. .

\end{document}